\documentclass[fleqn,usenatbib]{mnras}

\usepackage{newtxtext,newtxmath}

\usepackage[T1]{fontenc}
\usepackage{ae,aecompl}

\usepackage{graphicx}	
\usepackage{amsmath}	
\usepackage{xcolor}

\title[High resolution spectra of RR Lyrae]{Metallicities from high resolution spectra of 49 RR Lyrae Variables}

\author[Gilligan et al.]{
Christina K. Gilligan,$^{1}$\thanks{E-mail: christina.k.gilligan.gr@dartmouth.edu}
Brian Chaboyer,$^{1}$,
Massimo Marengo$^{2}$,  
Joseph P. Mullen$^{2}$,
\newauthor
Giuseppe Bono$^{3,4}$, 
Vittorio F. Braga$^{3,5}$, 
Juliana Crestani$^{3,4}$,
Massimo Dall'Ora$^{6}$,
\newauthor
Giuliana Fiorentino$^{3}$,
Matteo Monelli$^{7}$, 
Jill R. Neeley$^{8}$,
Michele Fabrizio$^{3,5}$,
\newauthor
Clara~E.\ Mart\'{i}nez-V\'{a}zquez$^{9}$, 
Fr\'{e}d\'{e}ric Th\'{e}venin$^{10}$,
and Christopher Sneden$^{11}$
\\
$^{1}$Department of Physics and Astronomy, Dartmouth College, Hanover, NH 03784, USA\\
$^{2}$Department of Physics and Astronomy, Iowa State University, Ames, IA 50011, USA \\
$^{3}$INAF-Osservatorio Astronomico di Roma, via Frascati 33, 00078 Monte Porzio Catone, Italy \\ 
$^{4}$Dipartimento di Fisica, Universit\`{a} di Roma Tor Vergata, via della Ricerca Scientifica 1, 00133 Roma, Italy\\
$^{5}$Space Science Data Center, via del Politecnico snc, 00133 Roma, Italy\\
$^{6}$INAF-Osservatorio Astronomico di Capodimonte, Salita Moiariello 16, 80131 Napoli, Italy\\
$^{7}$Instituto de Astrof\'{ı}sica de Canarias, Calle Via Lactea s/n, E38205 La Laguna, Tenerife, Spain\\
$^{8}$Department of Physics, Florida Atlantic University, 777 Glades Rd, Boca Raton, FL 33431 USA\\
$^{9}$Cerro Tololo Inter-American Observatory, NSF's NOIRLab, Casilla 603, La Serena, Chile\\
$^{10}$Universit\'{e} de Nice Sophia-antipolis, CNRS, Observatoire de la C\^{o}te d'Azur, Laboratoire Lagrange, BP 4229, F-06304 Nice, France\\
$^{11}$Department of Astronomy and McDonald Observatory, The University of Texas, Austin, TX 78712, USA\\
}

\date{Accepted XXX. Received YYY; in original form ZZZ}

\pubyear{2020}

\hypersetup{draft}
\begin{document}
\label{firstpage}
\pagerange{\pageref{firstpage}--\pageref{lastpage}}
\maketitle

\begin{abstract}
Accurate metallicities of RR Lyrae are extremely important in constraining period-luminosity-metallicity relationships ($PLZ$), particularly in the near-infrared.   We analyse 69 high-resolution spectra of Galactic RR Lyrae stars from the Southern African Large Telescope (SALT). We measure metallicities of 58 of these RR Lyrae stars with typical uncertainties of 0.15 dex. All but one  RR Lyrae in this sample has accurate ($\sigma_{\varpi}\lesssim10\%$) parallax from $Gaia$. Combining these new high resolution spectroscopic abundances with similar determinations  from the literature for 93 stars, we present new $PLZ$ relationships in WISE W1 and W2 magnitudes, and the Wesenheit magnitudes W(W1,V-W1) and W(W2,V-W2).
\end{abstract}

\begin{keywords}
stars: abundances -- stars: variables: RR Lyrae -- 
\end{keywords}



\section{Introduction} \label{sec:intro}

$H_{0}$ is one of the most important cosmological parameters in $\Lambda$CDM models. However, there is disagreement on the precise value of $H_{0}$. Measurements using early-time features (CMB, BAO, etc.) find a value of $H_{0}$=67.5$\pm$0.5 km s$^{-1}$ Mpc$^{-1}$ \citep{Planck}, while the most precise measurements using the cosmic distance ladder which is a local measurement of $H_{0}$, usually based upon Cepheids and SNeIa, find $H_{0}$ = 74.03 $\pm$ 1.42 km s$^{-1}$ Mpc$^{-1}$ \citep{Riess}. The difference in the early and late measurements of $H_{0}$ is over 4$\sigma$ and this difference is increasingly thought to not be due to systematics. In order to determine whether the cause is due to physics outside of $\Lambda$CDM rather than systematics in the measurements, different calibrators for the cosmic distance ladder need to be examined. We use the cosmic distance ladder to determine extragalactic distances.  First, distances to close objects are calibrated using geometrical methods and then used to calibrate farther objects, usually SNeIa.  This succession of methods is called the distance ladder. Since each subsequent rung of the distance ladder relies on the rungs that come before, changing the first rung distance of the distance ladder has the largest impact.

RR Lyrae can be used as one of the first rungs of the cosmic distance ladder to get an independent measure of $H_{0}$. Cepheids have been preferred since they are more massive and therefore more luminous than RR Lyrae. However, RR Lyrae stars have a number of advantages over Cepheids as distant indicators. As younger stars, Cepheids are not found in early-type galaxies, and are often  located in the actively star forming regions of galaxies, 
which can lead  to large extinctions and reddening. In contrast the older, lower mass RR Lyrae stars are present in all types of galaxies, and  can be found in regions with little or no extinction.

Another benefit of RR Lyrae is their lower luminosity dispersion as compared to Cepheids \citep{Bono03},  which implies that the uncertainties in the RR Lyrae period-luminosity-metallicity relationship will be smaller than those for Cepheids. We note that a period-luminosity-metallicity relationship for RR Lyrae only exists in the infrared bands. Finally, RR Lyrae stars are much more common than Cepheid stars.  However, RR Lyrae stars are still relatively rare, and until the advent of Gaia, there were few accurate parallaxes for them. \citet{Benedict} measured the parallaxes of 5 RR Lyrae using the Hubble Space Telescope (HST), the largest sample at that time. Fortunately, $Gaia$ has measured the parallaxes of thousands of RR Lyrae \citep{DR2} and even better parallaxes are expected soon with the third Data Release.

The most popular diagnostic to estimate the distances of individual RR Lyrae has been for decades the use of the visual magnitude-metallicity relation  ($M_V$ vs.\ [Fe/H]). However, this relation  is prone to several thorny problems. First, evolved RR Lyrae appear systematically brighter in the $M_V$,[Fe/H] plane and we do not have firm constraints on the evolutionary status of individual stars, as we lack solid estimates for the surface gravity of individual RR Lyrae stars.  Second, the $M_V$-[Fe/H] may not be linear over the entire metallicity range \citep[e.g.][]{Catelan, Marconi}. Third, the steep dependence of $M_V$ on [Fe/H] implies that uncertainties in the metallicities (either in individual estimates, or the adopted metallicity scale) has a significant impact on the derived distances. An error of $\sim 0.2-0.3$ dex in the metallicity estimate implies an error on the absolute visual magnitude of $\sim 0.06-0.10$ mag. Finally, reddening corrections are important in the visual; error of $\approx 0.02\,$mag in the reddening correction implies an error of $\approx 0.06\,$mag  in the derived distance modulus.

A significant fraction of the pitfalls affecting the visual magnitude-metallicity relation can be either avoided or limited when moving into the mid-infrared (3 - 5 micron) regime.  Dating back to \cite{Longmore}, it become clear that RR Lyrae obey  a well defined period-luminosity relation in the near infrared bands \citep[e.g., see][for a more recent discussion]{Braga}, and the same is true in the mid-infrared \citep[e.g.][]{Neeley2017, MuravevaGaia, Neeley2019}, There are two key advantages in using mid-infrared period-luminosity  relations: the slope becomes systematically steeper when moving from the I to the K-band and at the same time the standard deviation becomes smaller. Theoretical calculations indicate that the mid infrared period luminosity relations  are minimally affected by evolutionary effects and that they  are linear over the entire period range \citep{Neeley2017}. Moreover and even more importantly, fundamental and first overtone RR Lyraes obey independent period luminosity relations. Finally,  absorption in the mid-infrared is more thaan order of magnitude less than in the visual, with a corresponding decrease in the uncertainty related to absorption corrections.  Current theoretical \citep{Neeley2017} and empirical \citep{Neeley2019} evidence indicates that the mid infrared period luminosity relation depends on  on the metal content, but the coefficient of the metallicity term is typically smaller than $\sim 0.15$ dex.  In this context it is worth mentioning that individual distances based on Gaia DR2 trigonometric parallaxes agree quite well, within the errors, with theoretical predictions \citep{MuravevaGaia}.  Another motiviation to use period-luminosity relations in the mid infrared is that JWST/NIRCAM will have a high sensitivity at these wavelengths, leading to the possibility of obtaining RR Lyrae distances to a number of different galaxies.

Once the systematics of Gaia parallaxes are fully understood, the main source of uncertainty in using RR Lyrae as a standardizable candle will be the metallicity coefficient of the $PLZ$. Measuring metallicities of RR Lyrae is commonly performed using three methods: the $\Delta S$ method (from low resolution spectra), photometric metallicities through colours and shapes of the light curves,  and directly through high-resolution spectroscopy (HRS). The former two methods  typically rely upon the use of calibration stars with known metallicities to establish empirical relations that are used to convert a $\Delta S$ measurement, or light curve shape measurement into a metallicity. In contrast, HRS provides a direct measurement of the metallicity. 

The number of HRS metallicities of RR Lyrae from HRS has always trailed those of Cepheids and other variable stars. The periods of RR Lyrae are fairly short, almost always less than 0.9 days, making long exposures useless due to the cyclical velocity smearing. In addition, RR Lyrae stars span a wide range of metallicities, ranging  from $\sim-3.0$ \citep{Govea} to $\sim 0.1$ dex \citep{Chadid}, implying that a large variety of spectral lines need to be measured in order to find lines useful for metallicity determinations. 

At present, there are 105 RR Lyrae with metallicities determined from HRS \citep{Michele} and an additional 23 stars from \citet{Liu} with a [Fe/H] spread of $\sim$2.5 dex. This work aims to significantly expand the number of RR Lyrae with accurate metallicities by presenting 49 new metallicity measurements of RR Lyrae.

The long term goal is to create a large, homogeneous dataset to more precisely and accurately determine the near-infrared $PL_{J, H, K}Z$ relationships in order to independently determine $H_{0}$. We have also obtained at least 20 epochs of near-infrared (JHK) photometry for all of the stars examined in this work. \citet{Massimo} will analyse the photometry and present near-infrared lightcurves for these stars. 

We present our observations and data reduction in Section \ref{obs}. We then explain our stellar parameter fitting routine in Section \ref{analysis}. We compare the results we obtain with this routine for previously fit spectra in addition to comparing our metallicity results to previous measurements in Section \ref{results}. In Section \ref{PLZ}, we use these newly measured metallicities to create new $PL_{W1,W2,W}Z$ relationships. Section \ref{conclusion} contains our final conclusions and remarks.

\section{Observations and Data Reduction}\label{obs}

\subsection{Target Selection}

All of our target stars are present in the $Gaia$ DR2 catalog with positive parallaxes with uncertainties less than 10\%. We prioritised stars present in the \citet{Chadid} sample to be able to directly compare our work with previous studies. In addition, 7 stars were observed twice in order to validate the internal consistency of our analysis program. Priority is given to stars for which we have collected a full light curve in the near-infrared. The stars are listed in Table \ref{Observations}.


\subsection{Data}

Observations were performed with HRS on the Southern African Large Telescope (SALT). HRS is a dual beam (3700-5500 {\AA} and 5500-8900 {\AA}) echelle spectrograph. We used the medium resolution mode with $R \sim 37,000$. SALT is a queue observed telescope and allows for objects to be observed only at certain epochs if the ephemeris data are provided. Exposure times were chosen such that the S/N of the spectra was $\sim$100 accounting for various weather conditions and estimated luminosity based on pulsation period and phase. However, as discussed in Section \ref{analysis}, after recent refinement of the periods, we found that the phase of our observations was significantly different than what was initially thought so some spectra had worse S/N (with a lower limit of 40) and some had higher with an average S/N of 115. A number of studies have suggested that reliable [Fe/H] abundances are best determined in the range of phases $\sim 0.2 \sim  0.5$  \citep[e.g.][]{preston, for2011, Chadid} 
as in these phases it is thought that the RR Lyrae atmospheres are relatively quiet, without shocks or significant velocity gradients. Thus, the fact that our spectra were obtained at practical random phase is not ideal.  However, detailed studies have shown that the the derived [Fe/H] abundances do not show a significant dependence on the observed phase \citep[e.g.][]{for2011, Fossati}, and so we analyzed all of our spectra, regardless of the observed phase.  We return to this issue \S \ref{sect_synth} and \S \ref{sect_lit}.

\begin{table*}
	\centering
	\caption{Stars observed. Full table available as supplementary material.}
	\label{Observations}
	\begin{tabular}{lllllllll} 
		\hline
Name &$Gaia$ EDR3 ID &$\varpi$ [mas] &$\sigma_{\varpi}$ [mas] & Period [d] & $T_{0}^{*}$ [HJD] &Obs [HJD] &Phase &Type \\
\hline
AA Aql   &4224859720193721856 &  0.72056 &  0.01785 & 0.3617860175 & 2457194.01212 & 2458618.591308 & 0.63  & ab\\
AA CMi   &3111925220109675136 &  0.87177 &  0.01775 & 0.4763264561 & 2457305.072 & 	2458063.564907 & 0.38  & ab\\
AE Scl   &5027734380789950976 &  0.39706 &  0.02037 & 0.5501314496 & 2457277.87304 & 2458296.599236 & 0.22  & ab\\
AE Scl   &5027734380789950976 &  0.39706 &  0.02037 & 0.5501314496 & 2457277.87304 &   2458074.444962 & 0.79  & ab\\
AE Tuc   &4710156463040888192 &  0.57936 &  0.01405 & 0.4145285 & 2434273.232 & 2458275.647141 & 0.93  & ab\\
		\hline
\multicolumn{4}{l}{\textsuperscript{*}\footnotesize{$T_{0}$ is the time of minimum light.}}
	\end{tabular}
\end{table*}

\begin{figure*}
  \centering
  \begin{tabular}{c}
    \includegraphics[scale=0.7]{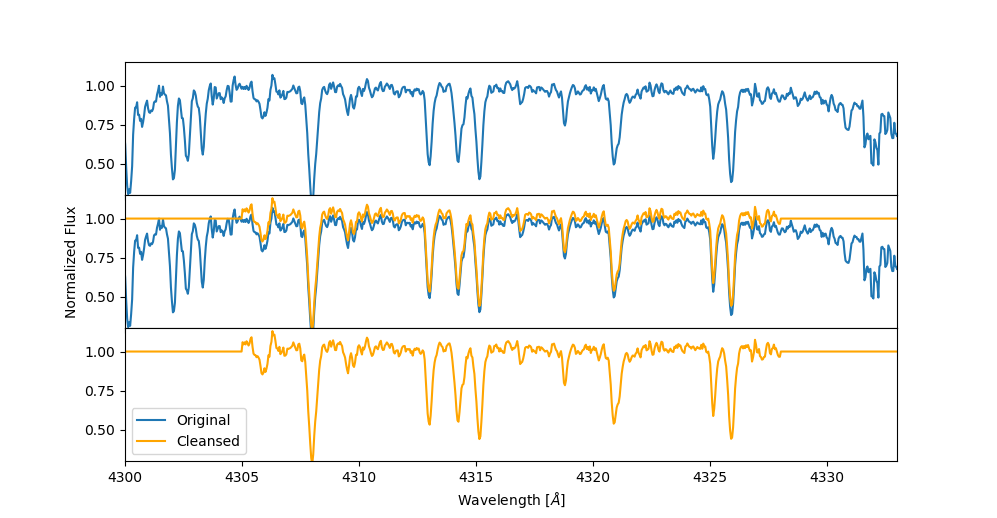}
      \end{tabular}
    \caption{Polynomial subtraction and masking for an example star, SX For. At the ends of the order, the MIDAS pipeline output is much noisier than the middle of the order. We are conservative in masking noisy regions, but changing the masked regions has very little impact in our metallicity determinations.}
    \label{fig:polynomialsubtraction}
\end{figure*}

The data are first reduced using the MIDAS \citep{Midas1, Midas2} pipeline, designed to be used for SALT spectra. MIDAS performs flat, arc, and object reductions, outputting a one-dimensional spectra. The MIDAS pipeline has some difficulties removing the echelle blaze pattern of the spectrograph, which is true for nearly all echelle pipelines. In order to lessen this effect, we fit a second order polynomial to each order of the spectrum, ignoring all features that are over 3 $\sigma$ from the fit. We then straighten the spectrum along this polynomial. In addition, there a number of areas of the spectra that are much noisier than the rest of the spectrum. We masked these bad areas of the spectrum. Luckily, these areas are fairly constant in placement on the CCD, so the same mask is applied to all of our spectra. This procedure is illustrated for a single echelle order in Figure \ref{fig:polynomialsubtraction}. Not including the masks does not greatly affect the results of our stellar parameter fitting program, but does add uncertainty to our fits. 

\section{Data Analysis}\label{analysis}

The [Fe/H] abundances are determined using a synthetic spectral analysis. This method takes a model atmosphere with set input stellar parameters along with a line list and outputs a synthetic spectra that we can then compare to our SALT spectra. We create synthetic spectra using the synth driver of the LTE code MOOG\footnote{The current version is available at \href{http://www.as.utexas.edu/$\sim$chris/moog.html}{http://www.as.utexas.edu/$\sim$chris/moog.html}} \citep{MOOG,NewMOOG}, which can then be compared to our observed spectra. 

The power of our fitting procedure is that we are able to fit the entire spectra at once, from around 4000 {\AA} to 7500 {\AA} using an extremely large grid of  stellar atmosphere models. The overall outline of our fitting procedure is as follows: (1) refine the line list; (2) create the grid of atmosphere models and synthetic spectra; (3) find the instantaneous stellar radial velocity and the spectral line Gaussian smoothing parameters; (4) calculate $\chi^{2}$ between the model and spectra; (5) recheck velocity and Gaussian smoothing parameters; and (6) if either the velocity or Gaussian smoothing parameters have changed, re-run comparison to models for models close in parameter space to the best-fitting model.

This automated method of determining [Fe/H] abundances was originally used as we had obtained a large number of spectra, many of which had lower signal to noise.  However, in the end our automated method did not converge for the lower signal to noise spectra, so the data presented in this paper all have  relatively high signal to noise ($\ga 50$).



\subsection{Line list creation}
\label{linelist}

The linelists are generated using the linemake code\footnote{Downloaded from \href{https://github.com/vmplacco/linemake}{https://github.com/vmplacco/linemake} and references therein} which includes both hydrides and molecules. In order to get the most accurate line list possible, every line in our wavelength region (4000 {\AA} -- 7500 {\AA}) has its oscillator strength ($\log{gf}$) adjusted to best match the spectrum of the Sun. Molecular lines play little role in determining the metallicity of these warm stars. 

\subsection{Creating atmosphere models \label{sect_synth} }

We used Kurucz stellar atmosphere models \citep{Kurucz}, with a fine grid of stellar parameters. Our choice of parameters are shown in Table \ref{Parmaeterspace}. In total, we create over 230,000 stellar atmosphere models, each of which is compared to all of our spectra.

\begin{table}
	\centering
	\caption{Parameter space of our Kurucz stellar models.}
	\label{Parmaeterspace}
	\begin{tabular}{lll} 
		\hline
Parameter & Range & Step \\
\hline
T [K] & 5000-8000 & 50\\
$\log g$ [dex] & 1.0-3.0 & 0.25\\
Microturbulent velocity ($\xi$)  [km s$^{-1}$]& 2.0-4.0 & 0.25 \\
$\text{[Fe/H]}$ [dex]& -3.00-0.00 & 0.05 \\
\hline
\end{tabular}
\end{table}

\subsection{Synthetic spectral analysis}

In order to directly compare our SALT spectra to the synthetic spectra created by MOOG, we need to shift our spectra in velocity space. In addition, we need to degrade our synthetic spectra by a Gaussian smoothing parameter to match the instrumental spectral resolution. Both of these parameters are initially determined at the beginning of the fitting procedure with an test synthetic spectrum that is in the middle of our parameter space. After an initial best-fitting set of models is found, the velocity and Gaussian smoothing parameters are re-fit to refine the validity of the initial values. 

\begin{figure*}
  \centering
\begin{tabular}{cc}
    \includegraphics[scale=0.5]{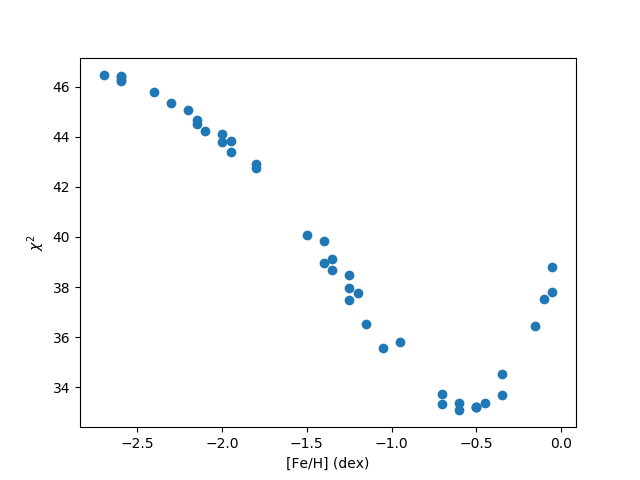}
     & 
    \includegraphics[scale=0.5]{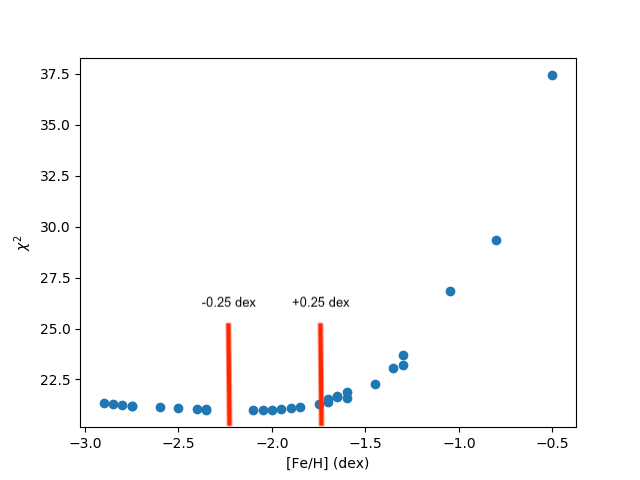}
\\
\small(a) & \small(b)\\
\end{tabular}
\caption{Two example fitting results. They are for the best-fitting temperature, $\log{g}$, and microturbulent velocity ($\xi$). Panel (a) shows the results for AA CMi. It is clear that our procedure is able to determine a best fit well. Panel (b) shows the results for AT Vir. It is clear that the fitting does not discriminate well between [Fe/H] values.}
\label{removed}
\end{figure*}

To find the best fit models, our program finds the minimum $\chi^{2}$ value between the grid of synthetic models and the SALT spectrum. 
For a number of stars, the $\chi^2$ values have a very shallow minimum, making it impossible to determine an accurate metallicity for these stars. These stars are removed from our sample of [Fe/H] measurements. 
Stars that are removed are ones  in which $\chi^{2}=1.1\chi^{2}_{min}$ has a [Fe/H] value that is more than 0.25 dex away from the best-fitting model's [Fe/H]. In essence, for these spectra, our fitting procedure does not discriminate between a wide range of [Fe/H] values. Figure \ref{removed}  shows an example for two different stars, AA CMi and AT Vir. AA CMi has a well-determined [Fe/H] but AT Vir does not. The 11 removed stars are marked with an asterisk in Table \ref{table:stellarparameters}. As discussed in \S  \ref{sect_lit}, the stars for which we are unable to determine a reliable [Fe/H] value were all observed at phases $> 0.5$, where the atmospheres are likely  more turbulent, and hence appear not to be well characterised by one-dimensional hydrostatic atmosphere models.

\subsection{Validation of our procedure}

To validate our procedure, we fit the reduced spectra from \citet{Chadid} with our automated program to determine [Fe/H], comparing the output stellar parameters with what was found by \citet{Chadid} who performed a synthetic spectral analysis by hand. These reduced spectra from \citet{Chadid} are entirely independent of our SALT spectra, and do not have stars in common.  Since many of the stars \citet{Chadid} were also examined in \citet{Layden}, we can directly compare our metallicity measurements to literature values from both \citet{Chadid} and \citet{Layden}. 

\begin{figure}
    \centering
    \includegraphics[scale=0.5]{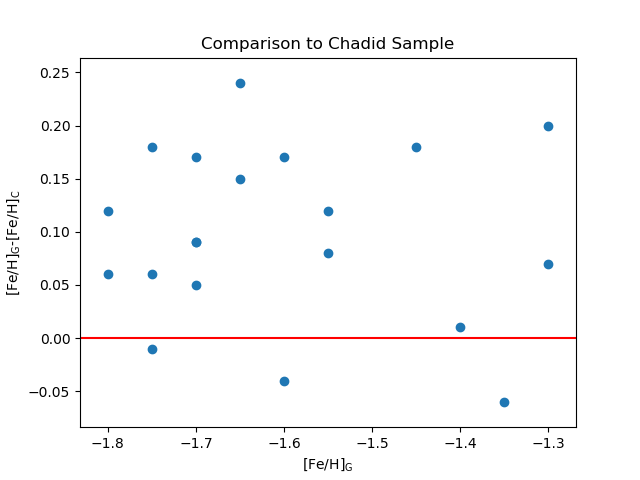}
    \caption{Comparison between the difference of [Fe/H] values found in this work $\left(\text{[Fe/H]}_{G}\right)$ and those found in \citet{Chadid} $\left(\text{[Fe/H]}_{C}\right)$.}
    \label{fig:chadid}
\end{figure}

Figure \ref{fig:chadid} compares [Fe/H] values found using our procedure ([Fe/H]$_{G}$) to the measurements from \citet{Chadid} ([Fe/H]$_{C}$). We removed stars that have poor metallicity determinations. After we remove these stars, we are left with 20 stars. The median difference is 0.09 dex with a standard deviation of 0.08 dex. There does not seem to be a trend with [Fe/H]. When directly comparing our best-fit stellar parameters to those found in \citet{Chadid}, our temperatures are nearly consistently 200-300 K hotter than in \citet{Chadid}, which at least partially explains the [Fe/H] offset since the two parameters are correlated. 


\begin{figure}
    \centering
    \includegraphics[scale=0.5]{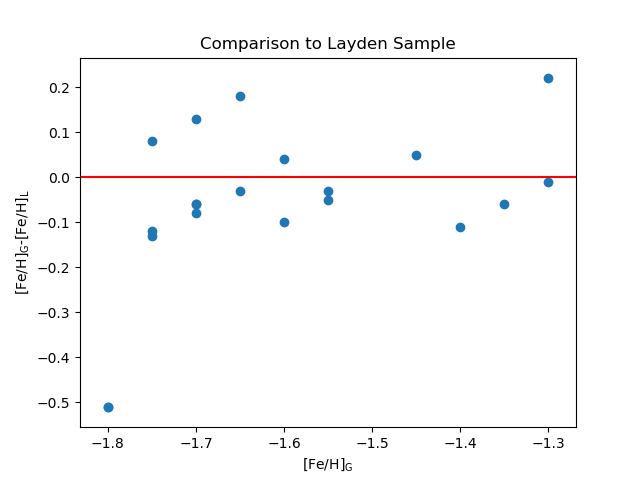}
    \caption{Comparison between the difference of [Fe/H] values found in this work $\left(\text{[Fe/H]}_{G}\right)$ and those found in \citet{Layden} $\left(\text{[Fe/H]}_{L}\right)$.}
    \label{fig:layden}
\end{figure}

Figure \ref{fig:layden} is similar to Figure \ref{fig:chadid} but instead compares our measured [Fe/H] ([Fe/H]$_{G}$) to the metallicities found in  \citet{Layden} ([Fe/H]$_{L}$). These are the same set of 20 stars in both Figure \ref{fig:chadid} and Figure \ref{fig:layden}. The median difference is -0.06 dex with a standard deviation of 0.18 dex, mainly due to the large outlier at [Fe/H]$_{G}$=-1.80 dex, VY Ser. \citet{Layden} found a [Fe/H]=-1.29 dex while \citet{Chadid} found a [Fe/H]=-1.86 for this star, closer to the value that we find in this work. If this outlier is removed from the comparison to the \citet{Layden} sample, the median difference is 0.04 dex with a standard deviation of 0.15 dex.

\section{Results}\label{results}

\begin{figure}
    \centering
    \includegraphics[scale=0.5]{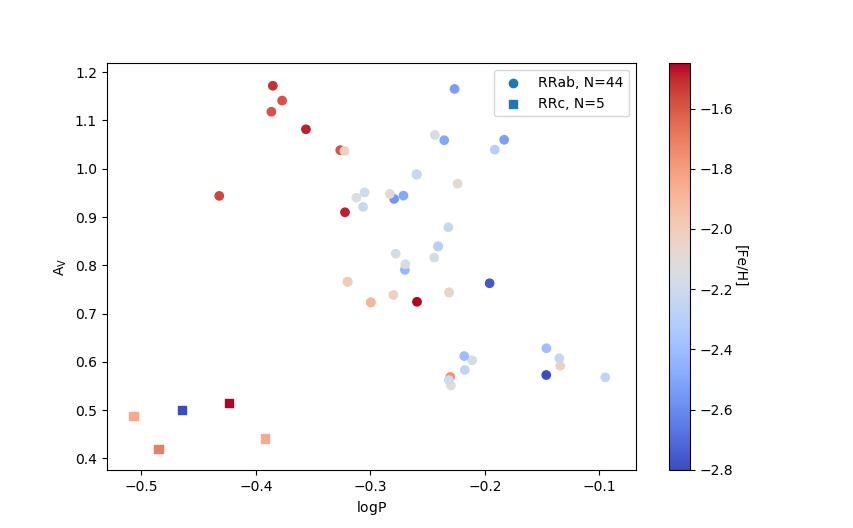}
    \caption{Bailey diagram for the stars for which we are able to find a best-fitting metallicity. The colors of the points correspond to the metallicity found in this work. The shapes indicate if the star is RRab or RRc.}
    \label{fig:bailey}
\end{figure}

The Bailey diagram for our is shown in Figure \ref{fig:bailey}. The RRc stars are well-distinguished from the RRab. Table \ref{table:stellarparameters} shows the best-fitting stellar parameters for our SALT spectra. It should be noted that varying the $\log{g}$ and $\xi$ parameters do not greatly affect the final metallicity determination. Hence, their values are not well-determined.

\begin{table}
	\centering
	\caption{Fitted stellar parameters of our RR Lyrae. Full table available as supplementary material. The estimated uncertainty in our [Fe/H] values is $\sigma_{\text{[Fe/H]}}$ = 0.15 dex.}
    \label{table:stellarparameters}
	\begin{tabular}{llllll} 
\hline
Name & [Fe/H] & Temperature [K] & $\log{g}$ [dex] &$\xi$ [km s$^{-1}$]\\
\hline
AA Aql$^{*}$ & -1.25 & 6400 & 2.75 & 3.00     \\
AA CMi & -0.6  & 6800 & 2.25 & 3.00     \\
AE Scl & -2.00    & 6700 & 2.75 & 3.00     \\
AE Scl & -1.70  & 6800 & 2.5  & 3.00     \\
AE Tuc$^{*}$ & -1.45 & 6800 & 2.75 & 3.00     \\
AF Vel & -1.60  & 6800 & 2.75 & 3.00     \\
\hline
\multicolumn{4}{l}{\textsuperscript{*}\footnotesize{Poor metallicity determination.}}
\end{tabular}
\end{table}

\subsection{Comparison to literature values \label{sect_lit}}

\citet{Michele} collected $\sim$2400 different metallicity measurements of RR Lyrae. Some of these stars ($\sim$300) only have metallicity measurements from \citet{Dambis}, which is based upon the catalog from \citet{Beers}, which mainly used HK objective prism spectra, or medium resolution  ($R\sim 4000$) spectra  to determine metallicities. A smaller sample of $\sim$100 stars instead have [Fe/H] abundances derived from HRS. For all other stars, \citet{Michele} derived metallicities using the $\Delta$S method for spectra with $R\sim$2000. Figure \ref{fig:michele} shows the comparison between our measured metallicities and those collected in \cite{Michele}.  The blue points only have measurements from \citet{Dambis}. The red points are RR Lyrae which have metallicity measurements from both \citet{Dambis} and high-resolution spectroscopy. The median differences and standard deviations between our [Fe/H] and those with previous HRS is -0.14 dex and 0.21 dex (10 stars), while for the entire sample is -0.12 dex and 0.41 dex (36 stars) respectively. If we remove the two largest outliers ([Fe/H]$\sim$ -1.2 and -0.2 dex), the median and standard deviations become -0.10 dex and 0.19 dex.

Figure \ref{fig:michelesnrphase} is similar to Figure \ref{fig:michele} but plot our residuals as a function of S/N and phase. There does not seem to be a correlation between our residual and S/N. However,  while we are always able to obtain a reliable [Fe/H] estimate when the phase is less than $\sim 0.5$, the fitting procedure often fails at higher phases, when the atmosphere is more turbulent.  Interesting enough, we are able to get reliable [Fe/H] abundances for many stars at phases $> 0.5$, though the dispersion between our [Fe/H] abundances and those found in the literature increases at the higher phases.

\begin{figure}
    \centering
    \includegraphics[scale=0.5]{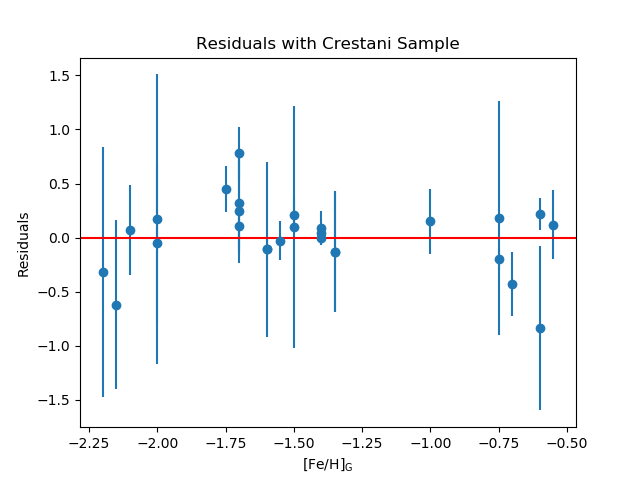}
    \caption{Comparison between the difference of [Fe/H] values found in this work $\left(\text{[Fe/H]}_{G}\right)$ and those found by \citet{Crestani} $\left(\text{[Fe/H]}_{c}\right)$. The error bars are only shown on the y-axis for clarity. There are 27 stars in common between the two independent analysis.}
    \label{fig:CrestaniDif}
\end{figure}

\begin{figure}
    \centering
    \includegraphics[scale=0.5]{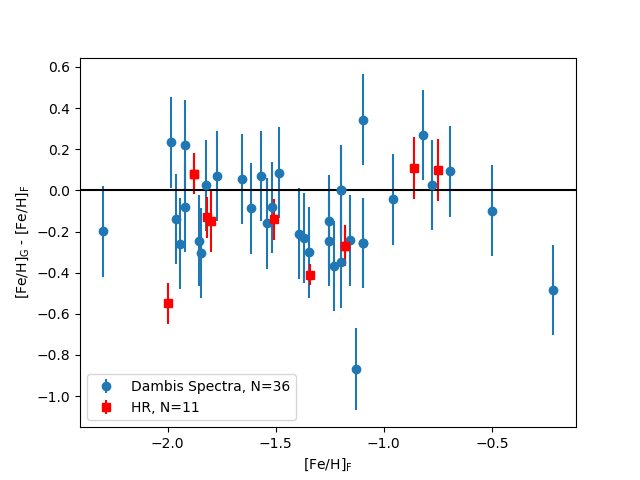}
    \caption{Comparison between the difference of [Fe/H] values found in this work $\left(\text{[Fe/H]}_{G}\right)$ and those found collected in \citet{Michele} $\left(\text{[Fe/H]}_{F}\right)$. The median difference between the two samples from HR is -0.14 dex while for the \citet{Dambis} spectra alone is -0.12 dex. The error bars are only shown on the y-axis for clarity.}
    \label{fig:michele}
\end{figure}

\begin{figure*}
    \centering
    \begin{tabular}{cc}
    \includegraphics[scale=0.5]{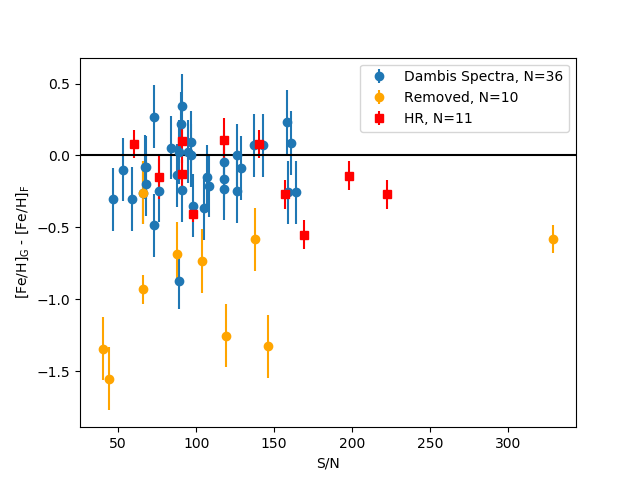}
    & \includegraphics[scale=0.5]{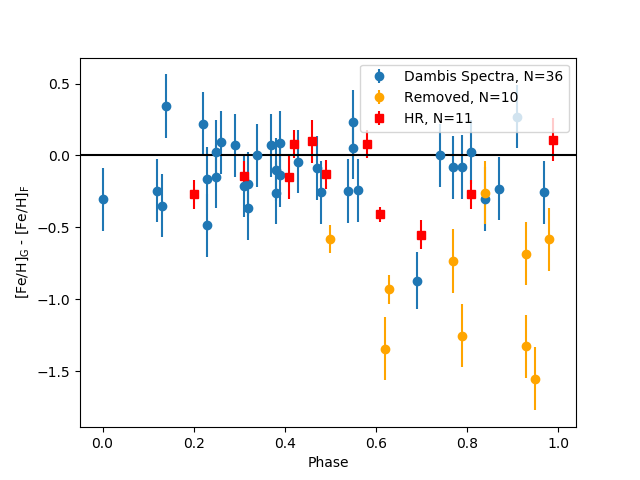}
    \end{tabular}
    \caption{Comparison between the difference of [Fe/H] values found in this work $\left(\text{[Fe/H]}_{G}\right)$ and those collected in \citet{Michele} $\left(\text{[Fe/H]}_{F}\right)$. The left panel compares the fits with S/N while the right compares with phase. The removed stars are ones where the fitting procedure is unable to distinguish a best-fitting metallicity. There does not seem to be a trend with respect to S/N. However, it is clear that closer to minimum light (phase = 1), it is more difficult for our routine to find a best-fit metallicity. For an RRab, the luminosity rises quickly after the minimum light which is likely why metallicities between a phase of 0-0.2 are easier to determine.}
    \label{fig:michelesnrphase}
\end{figure*}

The subset of the SALT spectra that contain well-observed Ca H and K lines are also analyzed in \citet{Crestani} who present a new calibration of the $\Delta S$ method for abundance determination. They perform a conventional  line-by-line abundance analysis that we can directly compare to our pipeline results since the spectra are the same. Figure \ref{fig:CrestaniDif} shows the residuals between \citet{Crestani} and our analysis. Overall, their is good agreement between the two analysis methods.  We also found that there was no correlation between the residuals and phase of observation.  We can use the differences between the [Fe/H] values  to estimate the error in our [Fe/H] determinations.  We calculated the root mean square between the two samples and then use the quoted errors from \citet{Crestani} to determine our error. The error in our [Fe/H] determination is 0.15 dex.

\begin{figure}
    \centering
    \includegraphics[scale=0.5]{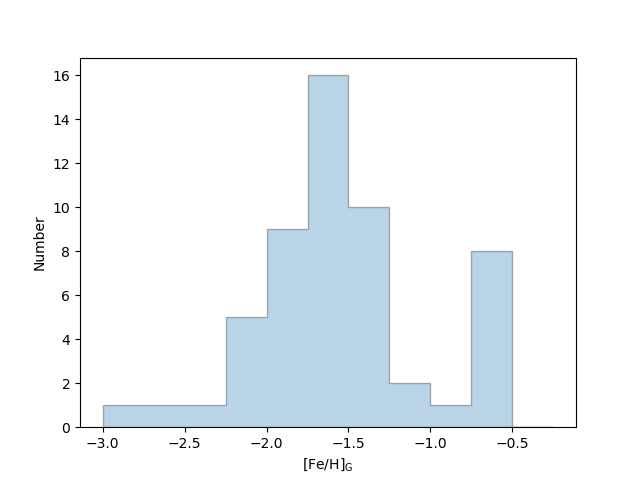}
    \caption{Histogram of our measured metallicities $\left(\text{[Fe/H]}_{G}\right)$. The shape is relatively Gaussian with a secondary peak of stars with an [Fe/H] between -0.75 to -0.5 dex. Due to the small number of stars examined and the fact that it is more difficult to measure metallicities for stars which have weak Fe lines, it is not unexpected to have this kind of population. In addition, since we chose our stars by apparent magnitude, we are biased toward RR Lyrae that are close to the Sun. Note that there is no physical reason why the metallicity distribution should be Gaussian. Apart from low number statistics and various biases, the distribution should reflect the overall [Fe/H] distribution in the halo.}
    \label{fig:hist}
\end{figure}

Figure \ref{fig:hist} shows a histogram of our measured metallicities. There is a Fe rich tail (AA CMi, AL CMi, SS Tau, U Pic, W Crt). Since we chose stars that were bright enough to be observed with short exposure times, we are biased toward stars close to the Sun and therefore of more similar Fe abundances to the Sun. In addition, it easier to determine metallicities for more metal-rich stars since the Fe lines are more apparent.

\section{Period-Luminosity-Metallicity Relationships}\label{PLZ}

Our newly determined metallicities were combined with high dispersion spectroscopic abundance determinations from the literature to come up with a sample of RR Lyr stars which have had their [Fe/H] values determined directly from high quality spectra.  Periods and mean magnitudes of this sample were taken from \citet{Mullen} to compute new $PLZ$ relationships. This paper utilises near-infrared photometry from both the Wide-field Infrared Survey Explorer (WISE) \citep{2010AJ....140.1868W} and its reactivation, the Near-Earth Object Wide-field Infrared Survey Explorer Reactivation Mission (NEOWISE) \citep{2011ApJ...731...53M} in the 3.4 and 4.6 $\mu$m bands, W1, and W2 respectively. Visible time-series photometric measurements, in the V band, were taken from the All-Sky Automated Survey for Supernovae (ASAS-SN), (\citet{2014ApJ...788...48S} and \citet{2018MNRAS.477.3145J}).
 
Periods are recalculated using the Lomb-Scargle method on both the V and W1 band data. The long survey length of the combined WISE and NEOWISE mission, running from late 2009-present, allows us to define a well-defined periods with accuracy on the order of $10^{-6}$ days or greater to properly phase without any drifting. Mean magnitudes for each band were determined by fitting the phased light curves with a Gaussian locally-weighted regression smoother (GLOESS) algorithm to gain a smoothed light curve. From the GLOESS light curves, characteristics such as amplitude, time of minimum light ($T_{0}$), and mean magnitude with its error can be easily calculated. For instance, the measurement of $T_{0}$ is defined as the epoch which lies closest in phase to the GLOESS smoothed light curve minimum. 

There are 120 stars that have data for both surveys and high  quality light curve fits. Out of these 120 stars, the max period difference between ASAS-SN and NEOWISE is  0.002 days, with 114 of these 120 star have period differences on the order of $10^{-5}$ days or smaller. The median period difference is $1.02x10^{-5}$ days, and we assume an uncertainty of $10^{-5}$ days in the period when performing our $PLZ$ fits.
For a more detailed description of the NEOWISE, ASAS-SN survey, light curve processing, various quality checks, and data extraction, we refer you to \citet{Mullen}. For specific reading on the GLOESS algorithm or the calculation of mean magnitude from a GLOESS light curve, please read \citealt{2004AJ....128.2239P} or \citealt{Neeley2015}, respectively.
 
The final sample contains 138 stars which have well determined periods, Gaia EDR3 parallaxes \citep{EDR3cat,EDR3pi}, W1 magnitudes and [Fe/H] values. To avoid biases which result from selecting a sample based upon parallax quality (e.g.\ removing negative parallaxes), the astrometric based luminosity (ABL) prescription \citep{ABL} was used to determine the  period-luminosity-metallicity ($PLZ$) relations for these stars.

\begin{table*}
\centering
\caption{Data used to derive our $PLZ$ and $PWZ$ fits. RRc stars have had their period fundamentalized by adding 0.127 to Log P. Parallaxes are from $Gaia$ EDR3 \citep{EDR3pi}; no zero point correction has been applied. Parallax uncertainties have been increased as disussed in the text.  Full table available as supplementary material.}
\label{PLZcatalog}
\scalebox{0.75}{
\begin{tabular}{lllllllllllll}
\hline
$Gaia$ EDR3 ID  & $\varpi$ [mas] & $\sigma_{\varpi}$ [mas] & W1 [mag] & $\sigma_{W1}$ [mag] & W2 [mag] & $\sigma_{W2}$ [mag] &  V [mag] &  $\sigma_{V}$ [mag] & Log P [days] & E(B-V) & [Fe/H] [dex] & $\sigma_{[Fe/H]}$ [dex] \\
\hline
3111925220109675136 & 0.8718 & 0.0248 & 10.2369 & 0.0064 & 10.2590 & 0.0065 & 11.5276 & 0.0252 & -0.322095 & 0.120 & -0.60 & 0.13\\
5360400630327427072 & 0.8308 & 0.0217 & 9.9791 & 0.0051 & 9.9927 & 0.0050 & 11.3856 & 0.0140 & -0.277834 & 0.319 & -1.60 & 0.13\\
3604450388616968576 & 0.6857 & 0.0359 & 10.0916 & 0.0057 & 10.1043 & 0.0055 & 11.4926 & 0.0096 & -0.211066 & 0.140 & -1.60 & 0.13\\
3677686044939929728 & 0.7579 & 0.0561 & 10.2009 & 0.0062 & 10.2126 & 0.0064 & 11.3491 & 0.0195 & -0.279211 & 0.112 & -2.20 & 0.13\\
\hline
	\end{tabular}}
\end{table*}

The PLZ relation was fit using the following equation
 \begin{equation}
\varpi 10^{0.2m_o - 2} = 10^{0.2 [a (\log P + 0.27) + b ([\mathrm{Fe/H}] + 1.3) + c ] }
 \label{eqnfitfunctionE}
\end{equation}
where $\varpi$ is the EDR3 parallax in mas, $m_o$ is the absorption corrected apparent magnitude, and the coefficients $a,b$ and $c$ are determined as part of the fit process. This explicit, non-linear fit, which takes into account the uncertainties in all of the observed quantities was performed using R \citep{rstats} and its nonlinear fitting function \texttt{nls}. 
In performing these fits, the uncertainty in the EDR3 parallaxes have been increased, based upon \S 7.1.2 of the EDR3 documentation \footnote{\url{https://gea.esac.esa.int/archive/documentation/GEDR3/Catalogue_consolidation/chap_cu9val/sec_cu9val_introduction/ssec_cu9val_intro_astro_precision.html}}  and Figure 19 in \cite{EDR3cat}.  In addition, an intrinsic dispersion of $0.03\,$mag was assumed to exist in the PLZ relation.

Considerable effort was put into reducing the systematic parallax  uncertainties in EDR3. However, the Gaia collaboration believes that there are still systematic parallax uncertainties which are a function of magnitude, colour, position on the sky and astrometric solution type, and  have published a calibration of this error \citep{EDR3bias}. We have applied this zero-point correction when performing the explicit PLZ fits.

EDR3 contains a number of  astrometric quality indicators, including the reduced uniform weight error, RUWE, and it is recommended that stars with RUWE $< 1.4$ be used when one desires to reduce the number of spurious parallaxes \citep{EDR3cat}.  From our original sample of 138 stars, 114 have RUWE $< 1.4$ and were used in an initial PLZ fit in the W1 filter.  This resulted in a very poor fit, mainly due to the inclusion of RR Lyr, which is a significant $4.5\,\sigma$ outlier from the fit.  To determine possible reasons why RR Lyr was an outlier, other astrometric goodness of fits indicators in EDR3 were examined: \texttt{astrometric\_excess\_noise, ipd\_gof\_harmonic\_amplitude, ipd\_frac\_multi\_peak} and \texttt{ipd\_frac\_odd\_win}. RR Lyr, along with a few other stars appears as an outlier in this sample, with relatively large \texttt{astrometric\_excess\_noise}.  In addition, some stars were outliers with relatively large values of  \texttt{ipd\_gof\_harmonic\_amplitude}, which is an indication of asymmetric images.   To remove these potential astrometric outliers, we required  \texttt{astrometric\_excess\_noise} $< 0.23$ and \texttt{ipd\_gof\_harmonic\_amplitude} $< 0.25$, which resulted in a sample of 108 stars which are used in the PLZ fits.  There are 5 RRc stars in this sample, with the rest of the sample being RRab stars. In performimg the PLZ fits, the periods of the RRc stars were fundamentalised by adding 0.127 to log P.  The relavent data for these 108 stars are given in Table \ref{PLZcatalog}.
 
 SALT [Fe/H] values were available for 36 stars, and the SALT [Fe/H] values were used in the fit when available. The remaining [Fe/H] values are from \citet{Michele}.  Since we found a $-0.14$ dex offset between our [Fe/H] values and the HRS [Fe/H] values collected in \citet{Michele}, an offset of $-0.14$ dex  was applied to the \citet{Michele} [Fe/H] values before the fit was performed.  The inclusion of this $-0.14\,$dex offset did not have a  significant impact on our fit results.
 
\begin{figure*}
 \includegraphics[scale=0.5]{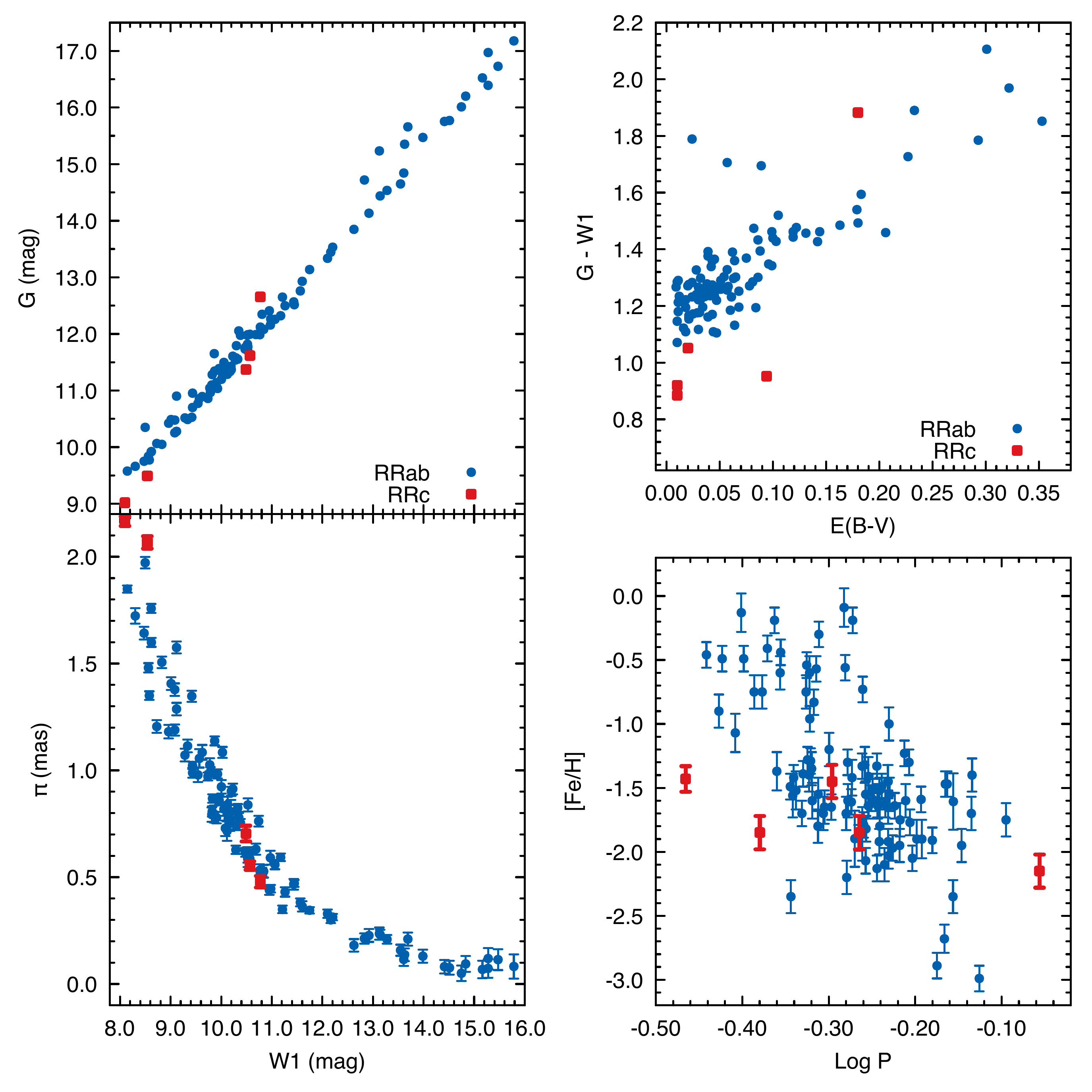}
 \caption{The photometric and parallax properties of the 108 stars used in the $PLZ$ fits (left panel), along with their colour--reddening,  log Period--[Fe/H] distributions (right panel). RRc stars have had their periods fundamentalised by  adding $0.127$ to $\log\,P_{FO}$. G magnitudes are from Gaia \citep{EDR3phot}.} 
 \label{fig_data2c}
 \end{figure*}

 Reddenings were determined from the 3-D maps from \citet{Green} as the first choice, or  \citet{Stilism}. As a test of the reddenings derived from the 3-D maps, these reddenings were compared to the 2-D reddenings from \citet{Schlegel}. The mean difference in E(B-V) is 0.03 mag with a median of 0.01 mag. In general, the 3D reddenings are less than the 2D reddenings when close to the galactic plane. There are only three stars near the galactic plane whose 3D reddenings are slightly larger than the 2D values, but the difference for all three stars is less than 0.01 mag in E(B-V). Conversion from E(B-V) to extinction  assumed $R_v = 3.1$,  $A_{W1}/A_v = 0.061$, and $A_{W2}/A_v = 0.048$. The overall properties of these stars used to determine PLZ relations are shown in Figures \ref{fig_data2c}. The reddening plot in the top right shows that most stars in this sample have relatively low reddenings, with  $E(B-V) < 0.2$, which corresponds to $A_{W1} < 0.038$.  Thus, uncertainties in the reddening values will not significantly impact the $PLZ$ fit. 
 
\begin{figure*}
 \begin{center}
 \includegraphics[scale=0.85]{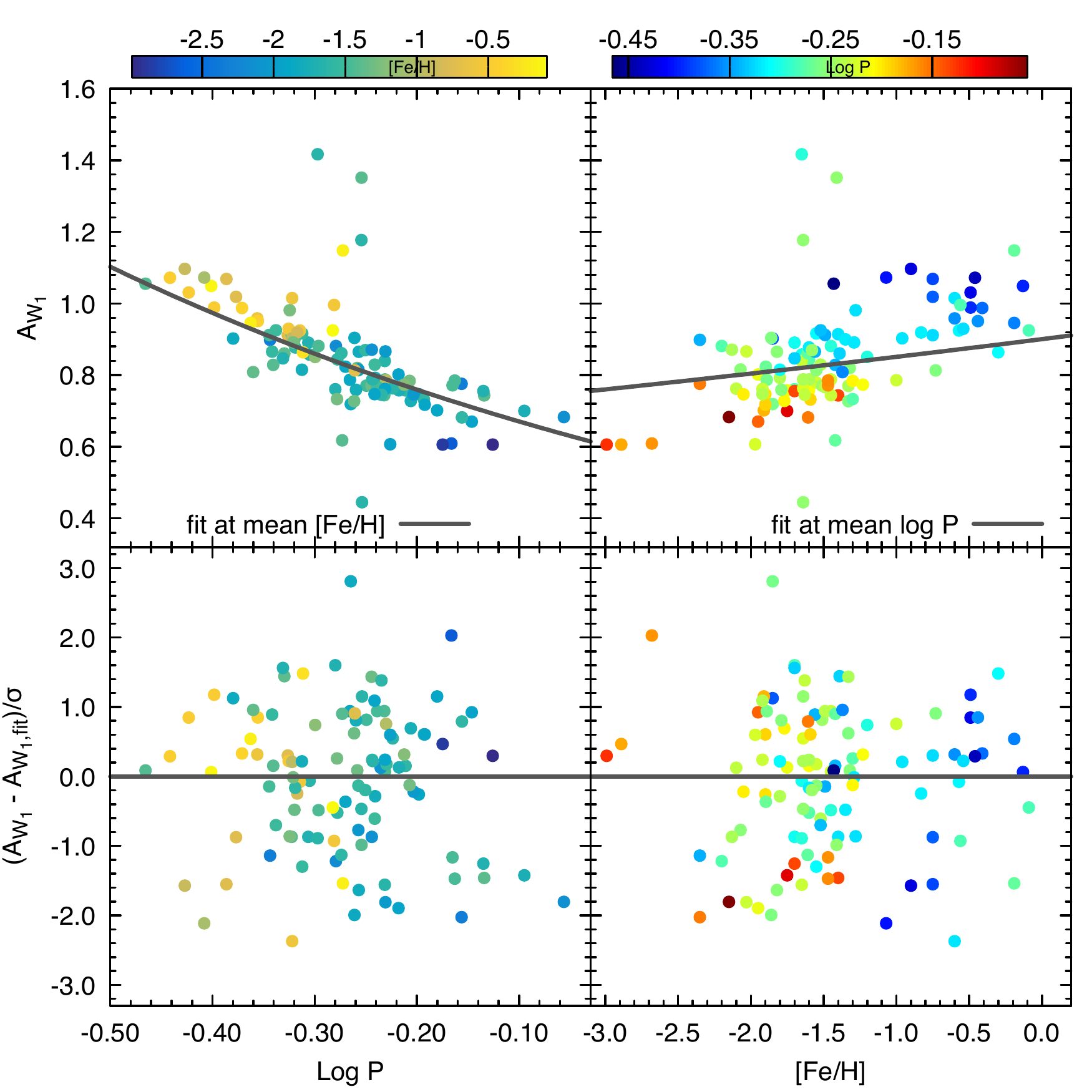}
\caption{The explicit P-W1-Z fit for the 108 star astrometrically clean sample. Note that $A_{W1}$ is not the absolute magnitude in the $W1$ filter, but comes from the ABL formalism and is defined as $A_{W1} =  = \varpi 10^{0.2 m_{W1,o} - 2}$. Uncertainties $A_{W1}$ (top panels) are not shown to make the figure clear, but can be inferred by looking at the normalized fit residuals which are shown in the bottom panels. Points are colored by their [Fe/H] values in the left panels, and their log P values in the right panels.}
 \label{fig_W1c}
 \end{center}
 \end{figure*}

  A $PLZ$ fit with the W1 magnitudes yields
\begin{eqnarray}
 \begin{aligned}[b]
 & M_{W1} = (-2.70\pm 0.12)(\log P + 0.27) \\
 &+ (0.123\pm 0.017)([\mathrm{Fe/H}] + 1.3)  - (0.390\pm 0.009)\; .
 \label{eqn_W1fit1}
  \end{aligned}
\end{eqnarray}
The quality of the fit is good, with a mean absolute normalized deviation of 0.82, where a value of 0.80 is expected for Gaussian uncertainties. The fit is shown visually in Figure \ref{fig_W1c}.

The explicit PLZ fit depends critically on the EDR3 parallaxes, and calibration of the Gaia EDR3 parallax zero-point error given by \cite{EDR3bias}. However \cite{EDR3bias} caution that their calibration is tentative and should be used with caution.  The primary  calibration of the zero-point error is based upon quasars which are numerous at fainter magnitudes, while at brighter magnitudes various secondary sources were used in the calibration.  In general, our stars are relatively bright, with 78\% of the sample having $G < 13$.  The catalogue validation carried out by the Gaia team \citep{EDR3cat} found that the parallax correction significantly improved the agreement with external data, except for the brighter stars (the LMC and SMC radial velocity samples which have median magnitudes of $<G> = 12.8$ and $<G>=12.5$ respectively).

To see what impact errors in the zero-point correction can have our our PLZ fit, we elected to perform an implicit fit, which solves for a global EDR3 parallax zero-point error as part of the fitting procedure: 
  \begin{eqnarray}
   \begin{aligned}[b]
  f & \equiv 10^{0.2 [a( [\mathrm{Fe/H}] + 1.3) + b(\log P +0.27) + c]} \\
   & - (\varpi + \beta) 10^{0.2m_{W1,o} - 2} = 0
   \end{aligned}
  \label{eqnfitfunctionI}
 \end{eqnarray}
where $\beta = \pi_{zp}$ is determined as part of the fitting process
Since our RR Lyrae sample is distributed randomly on the sky, and solving for a global zero-point error may be appropriate, even though EDR3 parallax zero-point error is known to vary spatially \citep{EDR3bias}.

A two step procedure was utilised when performing the fit.  First,  as in \cite{Layden19}, equation~\ref{eqnfitfunctionI} was fit using an a implicit, nonlinear, weighted orthogonal distance approach using ODRPACK95 \citep{Zwolak}.  This implicit fit accurately determines  the $PLZ$ zero-point and the DR2 global parallax offset, but Monte Carlo tests (see \citealt{Layden19}) indicated that the slopes returned by the implicit fit were biased. Hence, after performing the implicit fit, the derived EDR3 parallax zero-point error was adopted and used in an explicit nonlinear fit using R \citep{rstats} and its nonlinear fitting function \texttt{nls} to determine the slopes ($a$ and $b$ in Equation~\ref{eqnfitfunctionI}). Both of these fitting routines take into account the uncertainties in the observed quantities when determining the fit.

To verify the fitting procedure, five Monte Carlo tests were conducted. In this test, fake data was generated from a known $PLZ$ relation, where the distributions  (in period, [Fe/H], apparent magnitude, parallax uncertainties, etc) in the simulated data was drawn from the observed distributions.  The simulated data sets include the correlation between period and [Fe/H] which is present in our actual data. These five simulated data sets, with known $PLZ$ relations were then run through the exact same fitting routines as were used with our observed data. These simulated data sets had a  range of $PLZ$ slopes and zero-points, and a range of Gaia zero-point errors.  These simulations determined that the fitting routines are able to accurately recover the input $PLZ$ relation and Gaia zero-point error.  The average difference between the input coefficients and the fitted  coefficients was zero for the two zero-points and the [Fe/H] slope.   There is a slight tendency for the fitting routines to find a log Period slope which differs by $\sim 0.02$ from the true slope.  However, this difference is much smaller than the typical fit error ($\sim 0.16$) so we do not regard this disagreement as significant.  The standard deviation of the fitted parameters in the Monte Carlo tests was  very similar to the average error found by the fitting routines, indicating that the fitting routines are doing a reasonable job of estimating the uncertainty in the fitted coefficients.

An implicit P-W1-[Fe/H] fit which uses the zero-point corrected EDR3 parallaxes yields
   \begin{eqnarray}
   \begin{aligned}[b]
 & M_{W1} = (-2.77\pm 0.12)(\log P + 0.27) \\
 & + (0.119\pm 0.016)([\mathrm{Fe/H}] + 1.3)  - (0.421\pm 0.017)
     \end{aligned}
     \label{eqn_W1fit2}
 \end{eqnarray}
 with a global EDR3 parallax offset error of $+0.010\pm 0.007\,$mas. Compared to the explicit fit, the slopes are similar (with differences less than $1\,\sigma$) while the zero-point of the PLZ relation is fainter by 0.03 magnitudes, which is a $1.6\,\sigma$ difference. In these implicit fits, there is a strong correlation between the zero point in the PLZ relation and the Gaia EDR3 global parallax offset error. 
 
An  implicit P-W1-[Fe/H] fit which did NOT use the EDR3 parallax zero-point correction was also performed and found 
   \begin{eqnarray}
   \begin{aligned}[b]
 & M_{W1} = (-2.79\pm 0.11)(\log P + 0.27) \\
 & + (0.109\pm 0.016)([\mathrm{Fe/H}] + 1.3)  - (0.419\pm 0.016)
     \end{aligned}
     \label{eqn_W1fit3}
 \end{eqnarray}
with a global EDR3 parallax offset error of $-0.020\pm 0.006\,$mas.  It is interesting to note that the sign of the parallax offset error has changed.  These implicit determinations of the EDR3 parallax offset error  are consistent with the external catalogue validation done by the Gaia team, which found a negative parallax offset error prior to the zero-point correction for all samples \citep{EDR3cat}. The apparent magnitudes of our sample are most similar to the bright LMC star external catalogue, and \cite{EDR3cat} found a positive parallax offset error after the zero point correction was applied \citep[see Table 1 in ][]{EDR3cat}. The implicit PLZ fit finds also leads to a positive parallax offset error after the zero point correction is applied.  This suggests that the tentative  parallax offset correction given by \cite{EDR3bias} may be too large for our sample of bright stars. 

Looking at the global EDR3 parallax offset error that was found by the implicit fits with ($0.010\,$mas), and without ($-0.020\,$mas) the parallax zero point correction from \cite{EDR3bias}, one can estimate that if the zero point correction is multiplied by 0.65, then the implicit fits would not find a global EDR3 parallax offset error.  Accordingly,  the zero point correction for each star was multiplied by 0.65 and new fits were performed. As expected, the implicit fit yields a EDR3 parallax offset error of $0.000\pm 0.006\,$mas. The implicit fit yields a PLZ relation which is very consistent with the explicit fit  
   \begin{eqnarray}
   \begin{aligned}[b]
 & M_{W1} = (-2.78\pm 0.12)(\log P + 0.27) \\
 & + (0.115\pm 0.016)([\mathrm{Fe/H}] + 1.3)  - (0.417\pm 0.009)\, .
     \end{aligned}
     \label{eqn_W1fitbest}
 \end{eqnarray}
The quality of this fit is very good, with a a mean absolute normalized deviation of 0.81, an inspection of the residuals shows no trends, while a quantile-quantile plot indicates the residuals follow the expected Gaussian distribution.   
Given the tentative nature of the EDR3 parallax correction presented by \cite{EDR3bias}, the above fit is the most reliable  estimate of the P-W1-[Fe/H] relation for our data set.  However, the uncertainties in equation \ref{eqn_W1fit3} do not take into account the uncertainty associated with the EDR3 parallax zero point error and underestimates the possible error in the PLZ zero point. Given the different zero points found in the various fits (equations \ref{eqn_W1fit1} to \ref{eqn_W1fitbest}), our best estimate for the P-W1-[Fe/H] relation is given by equation \ref{eqn_W1fitbest}.

Using the same approach for the W2 magnitudes, one finds
    \begin{eqnarray}
   \begin{aligned}[b]
 & M_{W2} = (-2.80\pm 0.12)(\log P + 0.27) \\
 & + (0.117\pm 0.016)([\mathrm{Fe/H}] + 1.3)  - (0.402\pm 0.02) \,.
     \end{aligned}
     \label{eqn_W2fit}
 \end{eqnarray}
The quality of this fit is good, with a a mean absolute normalized deviation of 0.82,  no residual trends, and the residuals follow the   expected Gaussian distribution.

Wesenheit magnitudes are commonly used to derive RR Lyrae $PLZ$ relationships since these magnitudes do not rely on knowing the reddening of the individual stars \citep{Madore}. We define the Wesenheit magnitude using V and W1. We use the analytic form of the reddening law from \citet{Cardelli} and the extinctions for the WISE bands from \citet{Yuan}: 
 \begin{eqnarray*}
 W(W1,V-W1) = W1 - 0.065(V-W1)\, .
 \end{eqnarray*}
and we find
    \begin{eqnarray}
   \begin{aligned}[b]
 & M_{(W1,V-W1)} = (-2.83\pm 0.10)(\log P + 0.27) \\
 & + (0.123\pm 0.014)([\mathrm{Fe/H}] + 1.3)  - (0.492\pm 0.02)
     \end{aligned}
     \label{eqn_W2fit2}
 \end{eqnarray}
The quality of this fit is reasonable, with a a mean absolute normalized deviation of 0.85,  no residual trends, and the residuals follow the   expected Gaussian distribution.
We can also use W2 instead of W1 to form a Wesenheit magnitude with the following relation:
 \begin{eqnarray*}
 W(W2,V-W2) = W2 - 0.050(W2,V-W2)
 \end{eqnarray*}
and we find
    \begin{eqnarray}
   \begin{aligned}[b]
 & M_{(W2,V-W2)} = (-2.84\pm 0.10)(\log P + 0.27) \\
 & + (0.126\pm 0.014)([\mathrm{Fe/H}] + 1.3)  - (0.460\pm 0.02)
     \end{aligned}
     \label{eqn_W2fit3  }
 \end{eqnarray}
The quality of this fit is reasonable, with a a mean absolute normalized deviation of 0.87,  no residual trends, and the residuals follow the   expected Gaussian distribution. A summary of our PLZ relations is presented in Table \ref{tab_fits}.

 \begin{table*}
 \centering
 \caption{$PLZ$ and $PWZ$ relationship coefficients defined as $M=a(\log P + 0.27) +b([Fe/H]+1.3)+c$. $\alpha$ is the color coefficient used in defining a Wesenheit magnitude.\label{tab_fits}}
 \begin{tabular}{lcccc}
 Filter  & $\alpha^{*}$ & $a$ & $b$ & $c$ \\
 \hline
 \hline
 \multicolumn{5}{c}{$PLZ$}\\
 \hline
 W1 &  & $-2.78\pm0.12$ & $0.115\pm 0.016$ & $-0.417\pm0.020$ \\
 W2 &  & $-2.80\pm 0.12$ & $0.117 \pm 0.016$ & $-0.402 \pm 0.020$ \\
 \hline
 \multicolumn{5}{c}{$PWZ$}\\
 \hline
 W(W1, V-W1) & 0.065 & $-2.83\pm 0.10$ & $0.123 \pm  0.014$ & $-0.492 \pm 0.020$  \\
 W(W2, V-W2)& 0.050 & $-2.84\pm 0.10$ & $0.126 \pm 0.014$ & $-0.460 \pm 0.020$ \\
  \hline
 \multicolumn{4}{l}{\textsuperscript{*}\footnotesize{Only used for Wesenheit magnitudes.}}
 \end{tabular}
 \end{table*}

\subsection{Comparison to previous work}

\begin{figure*}
  \centering
  \begin{tabular}{cc}
    \includegraphics[scale=0.6]{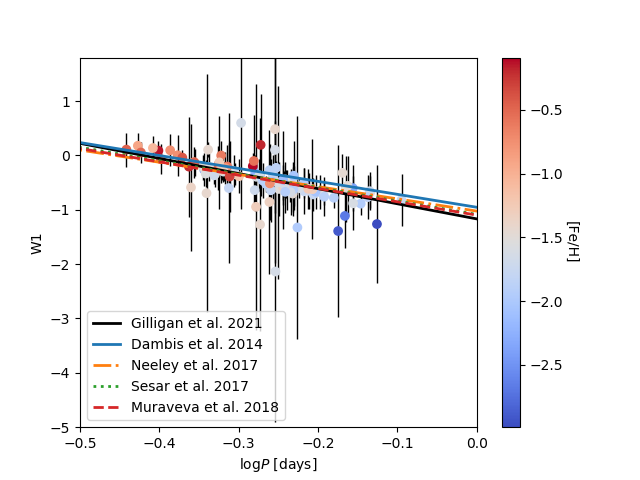} &  \includegraphics[scale=0.6]{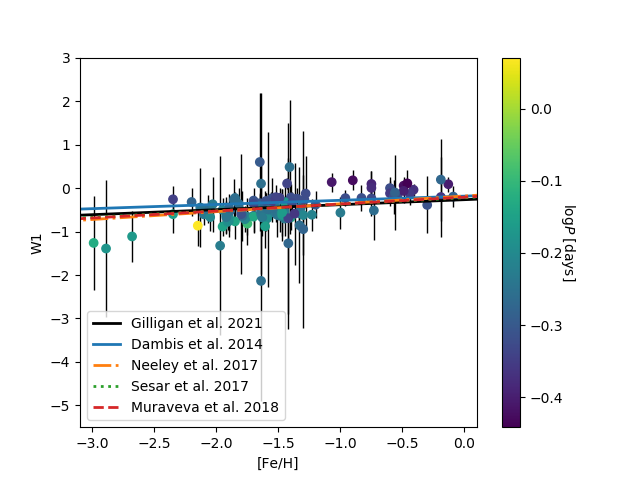}
      \end{tabular}
    \caption{P-W1-Z fits for the full HRS sample compared to the fits from \citet{Dambis2014}, \citet{Neeley2017}, \citet{Sesar2017}, and \citet{MuravevaGaia}. The left panel is the absolute magnitude of $W1$ versus $\log$P. The second panel is the absolute magnitude versus [Fe/H].}
    \label{pw1z}
\end{figure*}

\begin{figure*}
  \centering
  \begin{tabular}{cc}
    \includegraphics[scale=0.6]{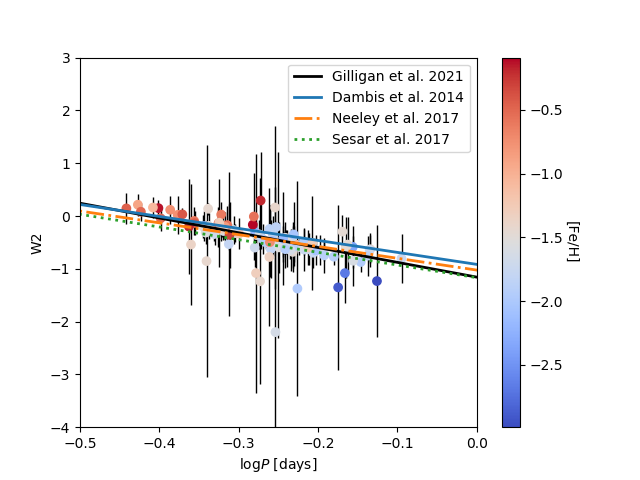} &  \includegraphics[scale=0.6]{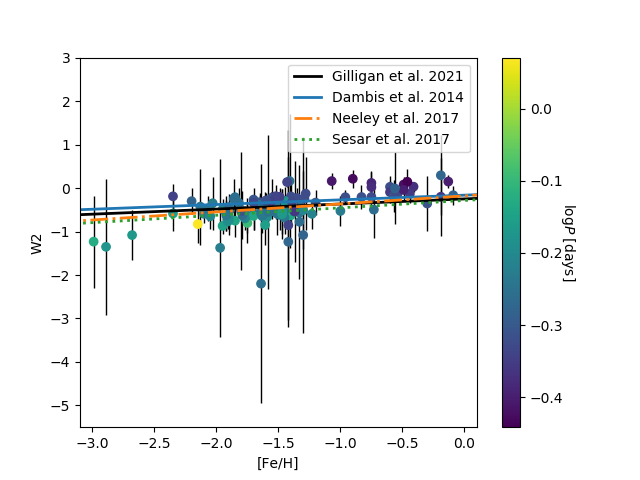}
      \end{tabular}
    \caption{P-W2-Z fits for the full HRS sample compared to the fits from \citet{Dambis2014}, \citet{Neeley2017}, and \citet{Sesar2017}. The left panel is the absolute magnitude of $W2$ versus $\log$P. The second panel is the absolute magnitude versus [Fe/H].}
    \label{pw2z}
\end{figure*}

A number of other studies have determined $PLZ$ relationships in infrared filters. Table \ref{PLZother} show a collection of previous $PLZ$ and $PWZ$ relations. Figure \ref{pw1z} shows our relationship plotted against each of the relations for the $W1$ band while Figure \ref{pw2z} shows the relations for the $W2$ band. It is clear that most of the differences between the relations occur with the more metal-poor and shorter period RR Lyrae. 


\citet{Dambis2014} collected WISE photometry from 15 Galactic globular clusters to use as calibrating stars. These clusters have known metallicities that are then applied to their member RR Lyrae. They also use \citet{Dambis} which uses a homogenized values of period, extinction, metallicity, and average magnitudes in optical and infrared passbands including $W1$. However, even though the catalog is homogenized, it is not homogeneous. The metallicity measurements come from both HRS and the $\Delta$S  method which is from low-resolution spectra. The zero-point ($c$) of the $PLZ$ is close to what is found in the other works and our zero-point. The period and metallicity slopes ($a$ and $b$) that \citet{Dambis2014} found are the shallowest of all the relations.

\citet{Neeley2017} created theoretical $PLZ$ relationships in a variety of optical, near-infrared, and mid-infrared filters using models of RR Lyrae. They use time-dependent convective hydrodynamical models with two different atmosphere models. Their models have a wide metallicity range ([Fe/H]$\sim$ -2.25 to +0.05 dex). The relations they derive are tested against Galactic and M4 RR Lyrae from \citet{Neeley2015} and find consistent distances derived from other techniques.

\citet{Sesar2017} used the Tycho-$Gaia$ Astrometric Solution (TGAS) along with WISE W1 and W2 measurements to create their $PLZ$ relations. This work uses photometry of 100 RRab stars (no RRc) in $W1$ and $W2$ along with Tycho-Gaia Astrometric Solution (TGAS) parallaxes. 

\citet{MuravevaGaia} used $Gaia$ DR2 parallaxes and single epoch WISE W1 data in their analysis of 397 stars. This work also uses the catalog from \citet{Dambis}, a non-homogeneous sample of RR Lyrae. There are 23 stars in their sample that have their metallicities derived from HRS. They determined the $Gaia$ DR2 global zero-point parallax error as part of their fitting procedure.

\citet{Neeley2019} used photometry from The Carnegie RR Lyrae program and parallaxes from $Gaia$ DR2 to determine $PLZ$ and $PWZ$ relationships to $\sim$50 Galactic RR Lyrae stars. They use HRS abundances from \citet{Fernley}. This is the only other sample that combines combines $Gaia$ DR2 parallaxes and homogeneous HRS abundances, although they did not determine  the $Gaia$ DR2 global zero-point parallax error as part of their fitting procedure.

These previous results typically included RRc stars in their analysis, with their periods fundamentalized in the same manner as was performed in our fits.  Compared to other studies our log period slope is steeper, while our slope with [Fe/H] is similar.  It is not clear what is causing this difference; it could be due to our use of HRS [Fe/H] values, our sample selection/size or the use of EDR3 parallaxes. We note that there are indications from theoretical models \citep{Neeley2017} that RRab and RRc stars follow different PLZ relations (even when the RRc stars have their periods funamentalized) and so the exact mix of RRab and RRc stars may impact the derived PLZ relation.   As other studies come out which use the EDR3 parallaxes, it will be interesting to see how their PLZ slopes compare to the relations derived in this paper.

A comparison of the PLZ  zero point found by different studies is summarized in Table \ref{zeropoint}.  Overall, there is reasonable agreement among the different zero points.  The uncertainty in our zero point is factor of 7 smaller than the  previous best observational determination by \cite{Dambis2014} and our zero point has the same uncertainty as found in theoretical models by \cite{Neeley2017}. Figures \ref{pw1z} and \ref{pw2z} show our fits plotted against various $PW1Z$ and $PW2Z$ relationships respectively. In the left panels, we plot with respect to $\log{P}$ (at fixed [Fe/H]$=-1.3$) while for the right panels we plot with respect to [Fe/H] (at fixed log P $=-0.27$).  One sees overall, a reasonable agreement between the fits, particularly when taking into account the relatively large zero-point uncertainties that are associated with previous observational determinations.

\begin{table*}
\centering
\caption{$PLZ$ and $PWZ$ relationship slopes defined as $M=a\log P +b[Fe/H]+c$.}
\label{PLZother}
\begin{tabular}{llccc}
Filter & Source & $a$ & $b$ & $c$\\
\hline
\hline
\multicolumn{5}{c}{$PLZ$}\\
\hline
W1 & \citet{Dambis2014} & -2.381$\pm$0.097 & 0.096$\pm$0.021 &-0.829$\pm$0.093 \\
 & \citet{Neeley2017} & -2.247$\pm$0.018 & 0.180$\pm$0.003 & -0.790$\pm$0.007\\
 & \citet{Sesar2017} & -2.470$^{+0.74}_{-0.73}$ & 0.150$^{+0.09}_{-0.08}$ &-0.890$^{+0.12}_{-0.10}$ \\
 & \citet{MuravevaGaia} & -2.450$^{+0.88}_{-0.82}$ & 0.16$\pm$0.10 & -0.910$^{+0.36}_{-0.34}$\\
$[3.6]$ & \citet{Neeley2019} & -2.40$\pm$0.27 & 0.18$\pm$0.03 & -0.793$\pm$0.007\\
W2 & \citet{Dambis2014} & -2.269$\pm$0.127 & 0.108$\pm$0.077 & -0.776$\pm$0.093\\
 & \citet{Neeley2017} & -2.237$\pm$0.018 & 0.185$\pm$0.003 & -0.785$\pm$0.007\\
 & \citet{Sesar2017} & -2.400$^{+0.84}_{-0.82}$ & 0.170$^{+0.10}_{-0.09}$ &-0.947$^{+0.11}_{-0.10}$\\
$[4.5]$ &  \citet{Neeley2019} & -2.45$\pm$0.28 & 0.18$\pm$0.03 & -0.785$\pm$0.007\\
\hline
\multicolumn{5}{c}{$PWZ$}\\
\hline
W([3.6], V-[3.6]) & \citet{Neeley2019} & -2.55$\pm$0.27& 0.18$\pm$0.03 & -0.46$\pm$0.02\\
\hline
\end{tabular}
\end{table*}

\begin{table*}
\centering
\caption{Comparison of zero-points. We set $\log{P}$=-0.27 and [Fe/H]=-1.3. For all of the previous studies, we are within 1$\sigma$.}
\label{zeropoint}
\begin{tabular}{llc}
Filter & Source & zero-point\\
\hline
\hline
\multicolumn{3}{c}{$PLZ$}\\
\hline
W1 & this work & -0.42$\pm$0.02 \\
& \citet{Dambis2014} & -0.31$\pm$0.14  \\
 & \citet{Neeley2017} & -0.42$\pm$0.02 \\
 & \citet{Sesar2017} & -0.47$\pm$0.80 \\
 & \citet{MuravevaGaia} & -0.44$\pm$0.37 \\
$[3.6]$ & \citet{Neeley2019} & -0.38$\pm$0.27 \\
W2 & this work & -0.40$\pm$0.02 \\
 & \citet{Dambis2014} & -0.30$\pm$0.18 \\
 & \citet{Neeley2017} & -0.42$\pm$0.02 \\
 & \citet{Sesar2017} & -0.52$\pm$0.84 \\
$[4.5]$ &  \citet{Neeley2019} & -0.36$\pm$0.28 \\
\hline
\end{tabular}
\end{table*}

\subsection{Testing the PLZ and PWZ relations}

To assess the implications of the PLZ and PWZ relations listed in Table~\ref{tab_fits} for distance determinations, we have used them to determine the distance to Reticulum and M4. These two old clusters  are optimal targets for testing these relations for two reasons. First, there are available data in the literature for their RR Lyrae stars in the $V$ band \citep{Kuehn2013, Stetson2014} and in the Spitzer IRAC 3.6 amd 4.5 $\mu$m filters, which are very similar to the $W1$ and $W2$ bands \citep{Neeley2015, Muraveva2018}. Second, these clusters contain 32 \citep{DemersKunkel1976, Walker1992} and 47 \citep{Clement2001,Stetson2014} RR Lyrae stars, respectively, therefore they are good candidates to test our relations with a statistical significance. To obtain accurate distance moduli, from the 30 RR Lyrae stars of Reticulum with $V, W1$ and $W2$ measurements, we reject the same stars that \citet{Muraveva2018} do not include in their calculations of the PL (i.e., V01, V08, V19, V24, V28, V32) since their position on the colour-magnitude diagram is unusual or they have noisy light curves.

In the case of M4, following \citet{Neeley2015} we reject two stars (V20 and V21) for being blends with nearby sources. Thus, we end up with a set of 31 RR Lyrae stars having $W1$ mean magnitudes and 28 with $W2$ mean magnitudes in M4. In order to obtain the true distance modulus ($\mu_0$) when using the PLZ relationships, we correct the $W1$ and $W2$ photometry by reddening. We consider $E(B-V) = 0.03$ \citep{Walker1992} for Reticulum and $E(B-V) = 0.37$ \citep{Hendricks2012} for M4. The extinction values for Reticulum ($R_V = 3.1$) are $A_{W1} = 0.203~E(B-V)$ and $A_{W2} = 0.156~E(B-V)$ \citep{Monson2012}. For M4, we calculate new absorption coefficients because its ratio of total to selective absorption is different ($R_V = 3.62$,  \citealt{Hendricks2012}). From \citet{Cardelli} we obtain $A_K/A_V = 0.124$ for $R_V = 3.62$ and using the $A_{W1}/A_K = 0.56$ and $A_{W2}/A_K = 0.43$ \citep{Indebetouw2005}, we get the the following extinction values, $A_{W1} = 0.251~E(B-V)$ and $A_{W2} = 0.193~E(B-V)$. It is worth noting that that the $\alpha^*$ coefficients for the Wesenheit magnitudes in M4 were also calculated accordingly to the latter extinction coefficients (i.e, $\alpha^*(W1,(V-W1) = 0.075$ and $\alpha^* (W2,(V-W2) = 0.056$). From our final selected sample of RR Lyrae stars, we obtain the distance moduli listed in Table~\ref{tab:distances}, adopting $\rm{[Fe/H]} = -1.66$ dex \citep{MackeyGilmore2004} as metallicity value for Reticulum and $\rm{[Fe/H]} = -1.10$ dex \citep[see][]{Braga2015} for M4.

\begin{table*}
\centering
\caption{True Distance moduli for Reticulum and M4 from our PLZ and PWZ relations (see Table~\ref{tab_fits}) and distance moduli from the literature}
\label{tab:distances}
\begin{tabular}{llcl}
\hline
\hline
Method & 
Filter & $\mu_0$ [mag] &  Reference \\
\hline
\multicolumn{4}{c}{Reticulum}\\
RR Lyr EDR3 calib. & PW(W1,V-W1) & $18.27\pm 0.10$ & this work\\
RR Lyr EDR3 calib. & PW(W2,V-W2) & $18.27\pm 0.10$ & this work\\
RR Lyr EDR3 calib.& W1 & $18.27\pm 0.11$ & this work\\
RR Lyr EDR3 calib.& W2 & $18.27\pm 0.13$ & this work\\
RR Lyr HST calib. & V & $18.40\pm 0.20$ & \cite{Kuehn2013}\\
RR Lyr Theoretical calib. &I & $18.47\pm 0.06$ & \cite{Kuehn2013}\\
RR Lyr  HST calib. &W1 & $18.43\pm 0.06$ & \cite{Muraveva2018} \\
RR Lyr TGAS calib. &W1 & $18.33\pm 0.06$ & \cite{Muraveva2018} \\
RR Lyr DR2 calib. &W1 & $18.32\pm 0.06$ & \cite{Muraveva2018} \\
RR Lyr HST calib. &W2 & $18.43\pm 0.08$ & \cite{Muraveva2018} \\
RR Lyr TGAS calib. &W2 & $18.34\pm 0.08$ & \cite{Muraveva2018}\\
RR Lyr DR2 calib.  &W2 & $18.34\pm 0.08$ & \cite{Muraveva2018}\\
RR Lyr Theoretical calib. &I & $18.51\pm 0.07$ & \cite{Braga2019}\\
RR Lyr Theoretical calib.&J & $18.47?\pm 0.10$ &\cite{Braga2019} \\
RR Lyr Theoretical calib. &K & $18.49?\pm 0.09$ &\cite{Braga2019}\\
RR Lyr Theoretical calib.&W1 & $18.30\pm 0.06$ &\cite{Braga2019}\\
RR Lyr Theoretical calib. &W2 & $18.31\pm 0.08$ & \cite{Braga2019}\\
RR Lyr Theoretical calib.&PW(V,B-I) & $18.52\pm 0.03$ & \cite{Braga2019}\\
\hline
\multicolumn{4}{c}{M4}\\
RR Lyr EDR3 calib.& PW(W1,V-W1) & $11.21\pm 0.08$ & this work\\
RR Lyr EDR3 calib.& PW(W2,V-W2) & $11.17\pm 0.08$ & this work\\
RR Lyr EDR3 calib. & W1 & $11.24\pm 0.09$ & this work\\
RR Lyr EDR3 calib. & W2 & $11.19\pm 0.09$ & this work\\
ZAHB & V &$11.28\pm 0.06$ & \cite{Hendricks2012}\\
Ecl.\ Binaries & ---&$11.30\pm 0.05$ & \cite{Kaluzny2013}\\
RR Lyr  HST calib. & W1 & $11.41\pm 0.08$ & \cite{Neeley2015}\\
RR Lyr  HST calib. & W2 & $11.39\pm 0.08$ & \cite{Neeley2015}\\
RR Lyr Theoretical calib. & R & $11.32\pm 0.11$& \cite{Braga2015}\\
RR Lyr Theoretical calib.& K &$11.30\pm 0.04$& \cite{Braga2015}\\
RR Lyr Theoretical calib.& PW(K,V-K) & $11.28\pm 0.05$ & \cite{Braga2015}\\
RR Lyr Theoretical calib.& R to K &$11.27\pm 0.02$ &\cite{Braga2015}\\
DR2 astrometry & ---& $11.38\pm 0.10$ & \cite{GaiaDR2glob} \\
Red Giant Oscillations & --- & $11.26\pm 0.06$ & \cite{Miglio2016} \\
DR2 astrometry & ---&$11.38\pm 0.10$ & \cite{Shao2019}\\
RR Lyr  DR2 calib. & I to W2 & $11.29\pm 0.02$ & \cite{Neeley2019}\\
\hline
\end{tabular}
\end{table*}


The systematic uncertainties in the distance moduli (shown in  Table~\ref{tab:distances}) are obtained by propagation of errors considering the photometric uncertainties of the mean magnitudes in $V$, $W1$ and $W2$, the uncertainties of the coefficients in the relationships (see Table~\ref{tab_fits}), and uncertainties of 0.2 dex in [Fe/H] and of 0.0001 days in the period (which are basically negligible in the final value of the uncertainties). In the case of the PLZ, we also take account of the uncertainty that comes from the reddening value, usually considered to be the 10\% of its value. The random uncertainty in our distance moduli, estimated from the standard error of the mean (the standard deviation divided by the square root of the number of RR Lyrae stars used to estimate the distance modulus) is quite small in all cases$\pm 0.01$ to $\pm 0.02\,$mag.

Overall, the distance moduli obtained with the different PLZ and PWZ relationships of Table~\ref{tab_fits} are identical for Reticulum and quite similar for M4.  Table~\ref{tab:distances} also contains a selection of distance estimates to these two clusters which have appeared in the literature since 2010. Most of these previous distance determinations are based upon RR Lyr stars, and these studies typically use the cluster RR Lyr stars  to find a log P slope for the PLZ relation, and a calibration of the PLZ zero point based upon theoretical models, or parallaxes (from either HST which obtained parallaxes of 5 stars, the Tycho-Gaia Astrometric Solution (TGAS), or $Gaia$ DR2). The first column in  Table~\ref{tab:distances} lists what method was used to determine the distance; if it based upon RR Lyr stars then how the zero point of the PLZ relation was calibrated is also listed. The uncertainties listed in Table~\ref{tab:distances} are those listed by the original authors, which in some cases (such as \cite{Muraveva2018}) are just the random uncertainties, and do not include uncertainties in the PLZ zero-point calibration.  \cite{Braga2015} provide a large number of distance estimates to M4 (all based upon a theoretical calibration of an RR Lyr PLZ) in a range filters, and only a small sample of their distance estimates are shown in Table~\ref{tab:distances}.

Modern estimates of the distance to Reticulum have all been obtained using RR Lyr stars.  The distances determined in this paper are lower than previous estimates, but agree within the uncertainties with  determinations which used mid-IR data (W1,W2).  The mid-IR distances are consistently smaller than those obtained at optical wavelengths, though the differences are not large.  For example our distance differs by $2.4\,\sigma$ from the largest distance in Table~\ref{tab:distances} from \cite{Braga2019} which is based upon a Wesenheit relation using V, B and I. We note that the distance determination depends somewhat on the assumed cluster reddening (for PLZ relations) and [Fe/H] value,  and different authors have used slightly different values, for example \cite{Braga2019} used [Fe/H]$=-1.70$.

Averaging together the distance determinations to the globular cluster M4 found in this paper leads to $\mu_o = 11.21\pm 0.09\,$mag. M4 has a relatively large ($\mathrm{E(B-V) = 0.37}$) and variable reddening with a non-standard extinction law \citep{Hendricks2012}.  The uncertainties introduced by this reddening are minimized in our mid-IR and Wesenheit distance determinations. In addition to RR Lyr based determinations, previous work has determined the  distance to M4 from theoretical zero-age horizontal branch (ZAHB) models, eclipsing binaries, solar-like oscillations in red giant stars and $Gaia$ DR2 astrometry of the brighter stars in M4.  The distances found in this work are once again shorter than previous distance determinations, though the differences are not large. In their paper on determining distances to globular clusters,  \cite{GaiaDR2glob}  note that there was a systematic difference of $-0.029\,$mas in their parallaxes as compared to the 2010 web update of the catalog of globular cluster properties  by  \cite{Harris1996}, and that the calibration noise is $0.025\,$mas. Due to other indications of a DR2 parallax offset error of $-0.029\,$mas, the \cite{GaiaDR2glob} distance to M4 shown in Table~\ref{tab:distances} has been corrected for this offset. The distance determined to M4 in this paper differs by $1.3 \sigma$ from those found by \cite{GaiaDR2glob}.  

Reasonable agreement is found with the distances determined with ZAHB models, eclipsing binaries and solar-like oscillations and many of the previous RR Lyr based determinations. For example,  our distance is  $0.9 \sigma$ smaller than the geometric eclipsing binary distance and $0.5 \sigma$ smaller than the distance determined from analysis of solar-like oscillations in red giant stars. 

Based upon the comparisons between distance estimates to M4 and Reticulum, it appears that our mid-IR calibration of the RR Lyr $PLZ$ relation, which is based upon $Gaia$ EDR3 parallaxes and high dispersion spectroscopic abundances, yields distances which agrees within the uncertainties with other distance determinations.  However, our distances are consistently smaller by $\sim 5 - 10\%$ compared to previous determinations.  It will be interesting to see how our distance scale compares to other EDR3 distance estimates in the future.

\section{Conclusion}\label{conclusion}

We performed a homogeneous spectral analysis of 49 RR Lyrae stars. This increases the number of RR Lyrae stars with HRS metallicity determination from 109 \citep{Michele} to 147. Our average error in our [Fe/H] determinations is 0.15 dex. We are able to compare our sample to results from three different methods, finding agreement between our overlapping stars. With these newly measured metallicities and data from \citet{Mullen}, we find new $PLZ$ relationships in the mid-IR using $Gaia$ EDR3 parallaxes.  These $PLZ$ relations are summarized in Table~\ref{tab_fits} and have substantially smaller uncertainties than previous observational determinations of the RR Lyr $PLZ$ relations in the mid-IR.  

This work will be extremely useful in \citet{Massimo} which will examine near-infrared (JHK) light curves for all of the RR Lyrae examined here and many more. The sample of stars with both accurate lightcurves and metallicity determinations will form the backbone of an independent $PLZ$ relationship that can be used for further rungs of the distance ladder.

\section{Acknowledgements}
The spectroscopic observations reported in this paper were obtained with the Southern African Large Telescope (SALT). This work has made use of data from the European Space Agency (ESA) mission {\it $Gaia$} (\url{https://www.cosmos.esa.int/gaia}), processed by the {\it $Gaia$} Data Processing and Analysis Consortium (DPAC,
\url{https://www.cosmos.esa.int/web/gaia/dpac/consortium}). Funding for the DPAC has been provided by national institutions, in particular the institutions participating in the {\it $Gaia$} Multilateral Agreement. This publication makes use of data products from WISE, which is a joint project of the University of California, Los Angeles, and the Jet Propulsion Laboratory (JPL)/California Institute of Technology (Caltech), funded by the National Aeronautics and Space Administration (NASA), and from NEOWISE, which is a JPL/Caltech project funded by NASA’s Planetary Science Division. MM and JPM were partially supported by the National Science Foundation under Grant No. AST-1714534. C. Sneden acknowledges support from NSF Grant AST1616040. 

\section{Data Availability}

The data (reduced spectra) can be found from SALT at \href{https://ssda.saao.ac.za/}{https://ssda.saao.ac.za/}. Other data underlying this article will be shared on reasonable request to the corresponding author.




\bibliographystyle{mnras}
\bibliography{rrlyrae} 





\bsp	
\label{lastpage}
\end{document}